# Large-area, Lithography-Free Super Absorbers and Color Filters at Visible Frequencies Using Ultrathin Metallic Films

Zhongyang Li [†], Serkan Butun [†], and Koray Aydin*

Department of Electrical Engineering and Computer Science, Northwestern University, Evanston, IL 60208, USA

* Email: aydin@northwestern.edu
[†] Authors equally contributed to this work.

**Nanostructured photonic materials enable control and manipulation of light at subwavelength scales and exhibit unique optical properties and photonic functionalities. In particular, plasmonic materials and metamaterials have been widely utilized to achieve spectral transmission, reflection and absorption filters based on localized or delocalized resonances arising from the interaction of photons with nanostructured materials. Realization of visible-frequency, high-performance, large-area, optical filters based on nanoplasmonic materials is rather challenging due to nanofabrication related problems (cost, fabrication imperfection, surface roughness) and optical losses of metals. Here, we propose and demonstrate large-area perfect absorbers and transmission filters that overcome difficulties associated with the nanofabrication using a lithography-free approach. We also utilize and benefit from the optical losses in metals in our optical filter designs. Our resonant optical filter design is based on a modified, asymmetric metal-insulator-metal (MIM) based Fabry-Perot cavity with plasmonic, lossy ultra-thin (~30 nm) metallic films used as the top metallic layer. We demonstrated a narrow bandwidth (~17 nm) super absorber with 97% maximum absorption with a performance comparable to nanostructure/nanoparticle-based super absorbers. We also investigate transmission (color) filters using ultra-thin metallic films, in which different**

**colors can be obtained by controlling the dielectric spacer thickness. With performance parameters of transmission peak intensity reaching 60% and a narrow-band of ~ 40 nm, our color filters exceed the performance of widely studied plasmonic nanohole array based color filters. Proposed asymmetric Fabry-Perot cavities using ultra-thin metallic films could find applications in spectrally selective optical (color and absorber) filters, optoelectronic devices with controlled bandwidth such as narrow-band photodetectors, and light-emitting devices.**

Plasmonic nanostructures and metamaterials provide unprecedented amount of opportunities in controlling and manipulating light-matter interactions at the nanoscale and enable spectral engineering of many optical processes including transmission, reflection, absorption and emission of light. In general, unique optical properties in such artificial engineered nanophotonic materials emerge from optical resonances induced by nanostructures, therefore spectral information depend strongly on the size, shape and periodicity of metallic and/or dielectric nanostructures. In recent years, plasmonic materials and metamaterials have received burgeoning amount of interest from the scientific community due to their exciting optical functions such as field localization, refractive index engineering, phase and amplitude control, and many more. Controlling the spectral transmission, reflection and absorption properties of materials is of special interest, in particular perfect absorbers and transmission (color) filters based on metallic nanostructures has been widely studied. It has been shown that almost perfect absorbers with zero reflection and transmission can be realized by designing metamaterials with an effective impedance matched to that of free space,[1] or fabricating metallic nanostructures on a dielectric/metallic substrate.[1-5] Since perfect absorbers based on plasmonic nanostructures and optical metamaterials operating at visible frequencies require subwavelength features, large-area and high-precision nano-fabrication techniques are required for functional visible-frequency perfect absorbers.

On the other hand, planar, multi-layer thin-film coatings can be designed as Bragg-reflector or Fabry-Perot cavities to yield resonant enhancement in various optoelectronic devices such as light emitting diodes (LED), photodetectors,[6-8] photovoltaics,[9, 10] surface photocatalysis, phototransistors,[11] modulators[12] and amplifier.[12, 13] Such planar optical devices can be traced back to a bulky, triple-layer structure known as Salisbury Screen (SS) which is designed for radar wave absorption.[13] More recently, several studies demonstrated strong resonant/absorptive characteristics from double-layer thin-film coatings[14, 15] or metal-insulator-metal triple-layer[16, 17] or multilayer[18, 19] coatings in microwave, infrared (IR) spectrum and beyond. Particularly, in a recent work[14] Kats *et al.* investigated strong optical interference effects in ultra-thin, absorptive Germanium films coated on metallic mirrors and demonstrated that continuous, absorptive thin-films with subwavelength thicknesses can indeed engineer the optical reflection and absorption spectra. This recent study renewed the interest in continuous, thin-film based optical devices and proved that ultra-thin continuous films are worth exploring as alternative candidates for spectral engineering of optical properties that are controlled solely by the film thickness rather than in-plane, nano/microstructure size. Continuous, unstructured thin-films involve only smooth planar surfaces, enabling high-throughput, large-area, lithography-free fabrication only by depositing thin-films on required substrates, therefore do not require costly nano-fabrication techniques

Here, we report large-area, lithography-free near-perfect (super) absorbers and transmission color filters based on triple-layer, metal-insulator-metal (MIM) thin-film stacks operating at visible frequencies. Super absorbers based on nanoparticles/nanostructures have previously shown to have broadband[3] or narrow-band[2, 4, 20-22] spectral characteristics. Resonant super absorbers are promising for various applications including photovoltaics, photodetectors, photothermal therapy,[23] thermophotovoltaics,[3] heat assisted magnetic recording,[24] hot-electron collection,[25] biochemical sensing[2] and thermal emitting.[1-5, 14, 15]

**Results**

Lossless dielectric materials with high index contrast have been widely used to support strong interference effects which depends on multiple pass light circulation in order to achieve anti-reflection/high-reflection coatings[26, 27] as well as optical filters.[28] For the energy harvesting/conversion purposes, materials with high dissipation (metals or semiconductors) need to be incorporated in the resonant cavity designs. However, lossy materials such as metals, reduce the quality factor of optical cavities by inducing attenuation and perturbation for propagation/trapped waves.[19, 29] Partially reflecting mirrors have been used as resonant

Fabry-Perot cavities for designing lasers, in which optical losses from lossy mirrors can be compensated by the gain media. Here, we present planar thin-film coatings comprised of MIM cavities, i.e., a thin metallic layer coupled through a dielectric spacer with the bottom metallic thick layer. The representative geometry is schematically shown in **Figure 1a** with size parameters of top Ag thickness $t$, middle $SiO_2$ thickness $d$ and bottom Ag thickness $h$. In this study, we will present two potential applications of such thin metallic multilayer films, namely super absorbers and color filters.

**Super Absorbers.**

For the design of super absorbers, we intentionally chose the bottom metal layer thickness to be optically thick ($h = 100$ nm) in order to minimize light transmission. To optimize the trade-off between light penetration into the planar nanocavity and material losses, top metallic film thickness $t$ is set to be 30 nm. This thickness allows for the balance between strong cavity confinement and light coupling (details are described in the following context and also Supplementary Information).

The simulated reflection spectra are obtained by performing full wave Finite-Difference Time-Domain (FDTD) method for the planar MIM cavity. The complex refractive index of Ag and $SiO_2$ is used from data of Palik.[30] A plane wave is normally incident to the top surface of MIM cavity. **Figure 1b** plots the reflection spectra as a function of $SiO_2$ thickness and the wavelength. As shown in Figure 1b, the reflection dip is shifting from ~ 400 nm continuously to ~ 750 nm as a function of the dielectric thickness $d$ from ~ 75 nm to ~ 200 nm, covering the entire visible frequency domain. A higher order mode starts to appear around 400 nm, when dielectric thickness $d$ is over 200 nm. In order to experimentally validate this effect, we deposited various multi-layer $Ag/SiO_2/Ag$ cavities with five different oxide thicknesses on a polished silicon wafer using electron-beam (E-beam) evaporation. The oxide thickness variations ($d = 90, 105, 130, 155$ and $175$ nm) are realized by etching the same oxide thickness coatings for different processing duration in Reactive-ion Etching (RIE) facility. Reflection measurements are performed using an inverted optical microscope setup. The simulated and measured reflection spectra for five different MIM cavities with varying oxide thicknesses are plotted in **Figure 1c** and **1d**. Our experimental and numerical results indicate that continuous three-layer MIM cavities enable strong optical interference effects and reduce reflection significantly from an otherwise conductive, highly reflective surface. The full width at half maximum (FWHM) for the measured resonances is ~ 14-28 nm, giving rise to quality factor (*Q*-factor) as high as ~ 48 ( $Q = \frac{\lambda}{\Delta\lambda}$ ).

Since the bottom metallic layer is optically thick, incident light is not transmitted through these films. Therefore, one can calculate the absorbance (*A*) of thin-film coating is using a simple formula *A = 1 – R - T (T = 0)*, where *R* and *T* represent the reflectance and transmittance, respectively. We plotted simulated (solid lines) and measured (dots) absorption spectra of five MIM films in **Figure 2a**. The overall shape and position of experimental spectral features match quite well with the simulated ones. The maximum absorption reaches over 99.5% in simulation and close to 96.8% in experiment. Specific performance parameters are listed in Table S1 in the Supplementary Information.

In order to visualize and better understand the absorption mechanism in three-layer MIM cavity, we calculated the total electric field (E-field) intensity (**Figure 2b**) and the absorbed power (**Figure 2c**) at the absorption resonance peak for the cavity with 155 nm thick oxide. The E-field is highly confined at the dielectric section between two metallic films, where a standing wave is formed due to constructive interference of incoming and reflected waves. Due to the enhanced electric field resulting from the cavity effect, most of the optical power is absorbed inside the top metallic film (Figure 2c). Absorption map reveals that absorption is higher at the metal-dielectric interfaces (both at the top as well as at the bottom interface). It is rather interesting to achieve almost perfect absorption using continuous metallic films. Essentially, the planar MIM cavity is an asymmetric Fabry-Perot-type (FP-type) resonator, comprised of lossless core dielectric with partially reflective top metal layer and optically thick, highly reflective mirror. The resonance condition is associated with the multiple round-trip phase shifts of electromagnetic wave inside the resonant cavity.[17] Nevertheless, owing to the dispersive nature of incorporated metal and its inherent resistive perturbation in films that is much thinner than the wavelength, the interface reflection and transmission phase changes along with the wave attenuation become non-trivial and render the resonant nanocavity studied here distinctive from conventional FP perspective.[19, 29, 31] Therefore, the top metallic thickness plays a critical role in achieving almost perfect absorption, which will be discussed later in detail.

In general, super absorbers operating at optical frequencies utilize nano and micro-structured surfaces that require nano-lithography and other processing techniques such as lift-off for patterning. In comparison with these nanostructure/nanoparticle-based absorbers [2-4, 21] the absorption bands in our proposed MIM planar cavities are quite sharp (~ 8-17 nm) with high, comparable peak absorption amplitude (~ 97%). Moreover, the absorption mechanisms in absorbers that utilize nanostructured materials usually depend on localized surface plasmon resonances (LSPR), which is highly susceptible to the inhomogeneity of features and surface roughness caused during the fabrication processes. MIM based absorbers proposed in this study involve sub-wavelength thickness films of metals and dielectrics that can be deposited using a single electron-beam or thermal deposition tool.

**Transmission Color Filters.**

As an alternative application of MIM films, we investigated transmission (color) filters at visible frequencies by reducing the thickness of bottom Ag film in order to allow light propagation. We explored five different MIM films with varying thicknesses *d*. Thicknesses of the top and bottom metallic films are chosen to 30 nm for optimum transmission filter performance, after initial electromagnetic simulations. Ag/SiO$_2$/Ag films are coated on double-side polished, transparent sapphire substrate. **Figure 3a** shows actual fabricated large-area color filters with five different oxide thicknesses. A white light source is excited from the back side of the filters and one can see vivid, bright colors of blue, cyan, green, orange and wine. Since the actual mechanism resulting in spectral selectivity in the transmission mode, is due to the constructive interference, the angle of incidence would affect the color. Angle-dependent characteristics of color filter are observed as shown in **Figure S2-S3** of Supplementary Information and more discussion on spectral shifting feature can be found correspondingly.

Measured and simulated transmission spectra for five MIM cavities with different oxide thicknesses are plotted in **Figure 3b**. An increase in thickness of dielectric layer $d$, results in red-shift of the cavity resonance wavelength. For the cases of MIM cavity with oxide thickness $d$ is ~ 100 nm, ~ 125 nm, ~ 150 nm, ~ 175 nm and ~ 200 nm, it is experimentally observed that five transmitted resonance peaks are equally distributed across the entire visible regime (Figure 3b). The maximum transmission peak obtained in measurement is ~ 61.6% and the narrowest bandwidth achieved is FWHM = 29 nm. Excellent agreement is obtained between the measured transmission spectra (dots) and the simulations (solid lines).

To prove the practical use of MIM transmission filters, we have patterned planar continuous MIM by etching the dielectric spacer layer using focused ion beam (FIB) milling. As shown in **Figure 4**, the transmission-mode micrographs of different pattern images are fabricated on double-side polished sapphire substrate by combining deposition of thin films and FIB approach.[14, 32] As an example, a defined concentric circle pattern (Figure 4a) was created with rainbow colors, in which the thickness of the dielectric layer is altered in a stepwise manner. Furthermore, a color palette consisting of 20 μm × 20 μm squares (Figure 4b) with different dielectric thickness $d$ was fabricated. Eventually, a tiny replica of 1932 oil painting *Le Rêve* (*The Dream*) by Pablo Picasso was fabricated on the resonant cavity by feeding the original image file to the FIB column control software (see Figure 4c). The software could calculate the etch depth as a function of the luminosity data in each pixel of the painting. The resulting pattern measures 178K dpi, which is roughly 500 times denser than any average smart phone display. For further comparison, the image shown here is about the size of a single pixel in an average smart phone display. Distinct from the common reflective imaging mode for human eyes, this technology by utilizing nanometre-thick films suggest the potential for transmission-mode design, labelling and visual arts of the future.[14]

For both absorption and transmission filters, top metallic silver film layer is chosen to be 30 nm. Note that this is a subwavelength thick metallic film for the operation wavelength range (400-800 nm). At visible frequencies metals do not behave like perfect dielectric conductors, therefore there is significant amount of light propagation inside the metal, rendering that metal film thickness is significantly important for the filter performance. This is quite different from conventional Fabry-Perot cavities, in which the thickness of partially reflective mirror is not taken into account. Due to the fact that the metal film used in the MIM filter design has plasmonic behavior at this wavelength, the effective cavity length is modified with the thickness of the metal films.

In order to investigate the effect of metal thickness on the optical performance of absorption and transmission filters, we performed additional numerical simulations and optical measurements. **Figure 5a** plots the absorption spectra of an MIM absorption filter with dielectric thickness, $d = 160$ nm and bottom Ag thickness $h$ is ~ 100 nm) as a function of the top metal film thickness. **Figure 5b** and **5c** show the measured and simulated absorption spectra for top metal thicknesses, $t = 15$ nm, 30 nm and 45 nm. When increasing the thickness of top metal $t$ between 15 nm and 45 nm, the bandwidth (FWHM) of absorbance is sharpened remarkably from ~ 51 nm to ~ 8 nm. Correspondingly, the thicker metallic boundary serves to boost the $Q$-factor of nanocavity from ~ 10 to ~ 75 and contributing to enhanced confinement of trapped wave and stronger wavelength-selective capability. More interestingly, as increasing top metal thickness $t$, the spectral position is not red-shifted, instead blue-shifted slightly from 645 nm to 623 nm. To clarify the counter-intuitive blue-shift, more in-depth discussion of analytical modelling to address the function of top metal layer can be found in **Figure S4** of Supplementary Information.

In addition to bandwidth control and resonance position modification, there also exhibits an optimum absorptivity when the top Ag thickness $t$ is around 30 nm. To achieve maximum absorption performance, there exists a trade-off between light penetration/coupling into cavity and the metallic losses by amount of the metal used in the filter design. Specifically, a thicker reflective metallic film result in higher reflection from the metal-air interface and will allow less light propagation through the dielectric layer. On the other hand, a thinner metallic film will result in higher transmission and leakage from the MIM cavity after reflecting from the bottom metallic mirror. For the case of $t = \sim 30$ nm, the maximum absorption achieved is attributed to optimization the interplay between cavity confinement and thin-film lossy nature.

Similar spectral evolutions are observed for MIM transmission filters upon changing the top boundary Ag layers while keeping the thickness of the middle $SiO_2$ layer at $d = 175$ nm and the bottom Ag layer at $h = 30$ nm, as displayed in **Figure 5d-5f**. Analogous spectral variations to perfect absorbers were exhibited as a function of top Ag layer thickness. In comparison with the reflection-mode cavity with optically thick metal, the spectra for leaking-mode cavity exhibits slightly broader bandwidth due to the weaker confinement from thinner reflective metallic films. To gain more insight into the impact of metal's optical property on absorption performance, we have also studied different metals such as Au/Ag and their hybrid combination for cavity and more discussion can be found in **Figure S5** of Supplementary Information.

**Discussion**

In conclusion, we have systematically investigated optical properties of asymmetric, modified Fabry-Perot-type cavity using large-area, lithography-free planar MIM coatings, which can be widely applied and integrated to optoelectronic devices including perfect absorbers and color filters at visible and near-IR wavelength range. We proposed and demonstrated a super absorber with a maximum absorption intensity of ~ 97% and a FWHM of 8 nm, whose performance is comparable with nanostructured plasmonic material based super absorbers. The thickness and the type of the metal film used in MIM cavity design significantly impacts the bandwidth, absorption and transmission intensity, the resonance wavelength and the quality factor of the FP cavities. By optimizing the interplay between the lossy nature of metal and the sub-wavelength thickness of the metallic films, the FP-type cavity enables high optical confinement of electric field in the middle dielectric layer and enhances optical absorption. By constructing leaking-mode nanocavity, transmission color filters are achieved with over 60% transmittance across the entire visible range and beyond. The high transmission efficiency and ability to realize of wide range of colors by the planar smooth surfaces is quite promising for potential filter and display devices integrated in future commercialized color-CCD/camera. Distinctive from the conventional reflective images for human eyes, this technology shows promise for transmission-mode design, coloring/labelling and visual arts of the future. The lithography-free approach based on ultra-thin metallic films or potentially 2D atomic materials [8, 11, 33, 34] provides great flexibility and high-throughput manufacturing convenience and open route for low-cost and robust optoelectronic devices based on ultra-thin film optical filters.

**Experimental and Simulation Section**

For perfect absorber samples, the bottom Ag layers of ~100 nm thickness were coated on polished silicon wafer using electron beam (E-beam) evaporator. For transmission filter samples, the thin (30 nm thickness) bottom Ag layers were coated on a double side polished sapphire substrate. Same thickness (~200 nm) of the $SiO_2$ layer was deposited onto the bottom Ag layer by E-beam evaporator. The different oxide thicknesses are processed by Reactive Ion Etching (RIE) facility to etch the same $SiO_2$ thickness coatings for different duration of time. The actual etching rate is ~ 33 nm/minute. Then 15-45 nm thick top Ag layers were deposited on the surface of $SiO_2$/Ag coatings by electron-beam evaporator. The micrographs shown in Figure 4 are fabricated through (1) depositing the bottom Ag and $SiO_2$ layers and (2) etching the patterns on the coating surface using Focus ion beam (FIB) milling and then (3) depositing the top Ag film. By feeding the original image files to the FIB column control software, it could calculate the etching depth as a function of the luminosity data in each pixel of the images.

A microscope equipped with a spectrometer consisting of a 303-mm-focal- length monochromator and Andor Newton electron multiplication charge-coupled device (EM-CCD) camera was utilized for optical characterization. Broad band illumination is generated by a broadband halogen lamp and a linear polarizer was inserted into the light pathway to polarize the incident light. Reflected and transmitted light was collected using a 2X Nikon microscope objective with a numerical aperture of 0.06. For calibration of reflection, we first measured the reflection from a broadband dielectric mirror (Edmund Optics #64-114) with an average reflection of 99% between 350 and 1100 nm. Measured reflection from MIM samples was then calibrated using the reflection spectra of the dielectric mirror.

Full-field electromagnetic wave calculations were performed using Lumerical, a commercially available finite-difference time-domain (FDTD) simulation software package.

Simulations for the planar MIM films were performed in 2D layout. A unit cell of 200 nm along the *x*-axis was selected for the planar structure and was simulated using periodic boundary conditions along the *x*-axis and perfectly matched layers (PML) along the propagation of electromagnetic waves (*y*-axis). Plane waves were launched incident to the unit cell along the +*y* direction, and reflection is collected with a power monitor placed behind the radiation source; transmission is collected with a power monitor placed behind the structure. Electric and magnetic field distribution cross-section are detected by a 2D field profile monitors in x-y plane. The complex refractive index of Ag for simulation is utilized from the data of Palik (0-2 μm) [30] and SiO2 is from the data of Palik.[30]

**Supporting Information**

Supporting Information is available from the Wiley Online Library or from the author.

**Acknowledgements**

This material is based upon work supported by the AFOSR under Award No. FA9550-12-1-0280. K.A. acknowledges financial support from the McCormick School of Engineering and Applied Sciences at Northwestern University and partial support from the Institute for Sustainability and Energy at Northwestern (ISEN) through ISEN Equipment and Booster Awards. This research was also partially supported by the Materials Research Science and Engineering Center (NSF-MRSEC) (DMR-1121262) of Northwestern University. This research made use of the NUANCE Center at Northwestern University, which is supported by NSF-NSEC, NSF-MRSEC, Keck Foundation, and the State of Illinois and the NUFAB cleanroom facility at Northwestern University. Z.L. gratefully acknowledges support from the Ryan Fellowship and the Northwestern University International Institute for Nanotechnology.

**Figures and Captions**

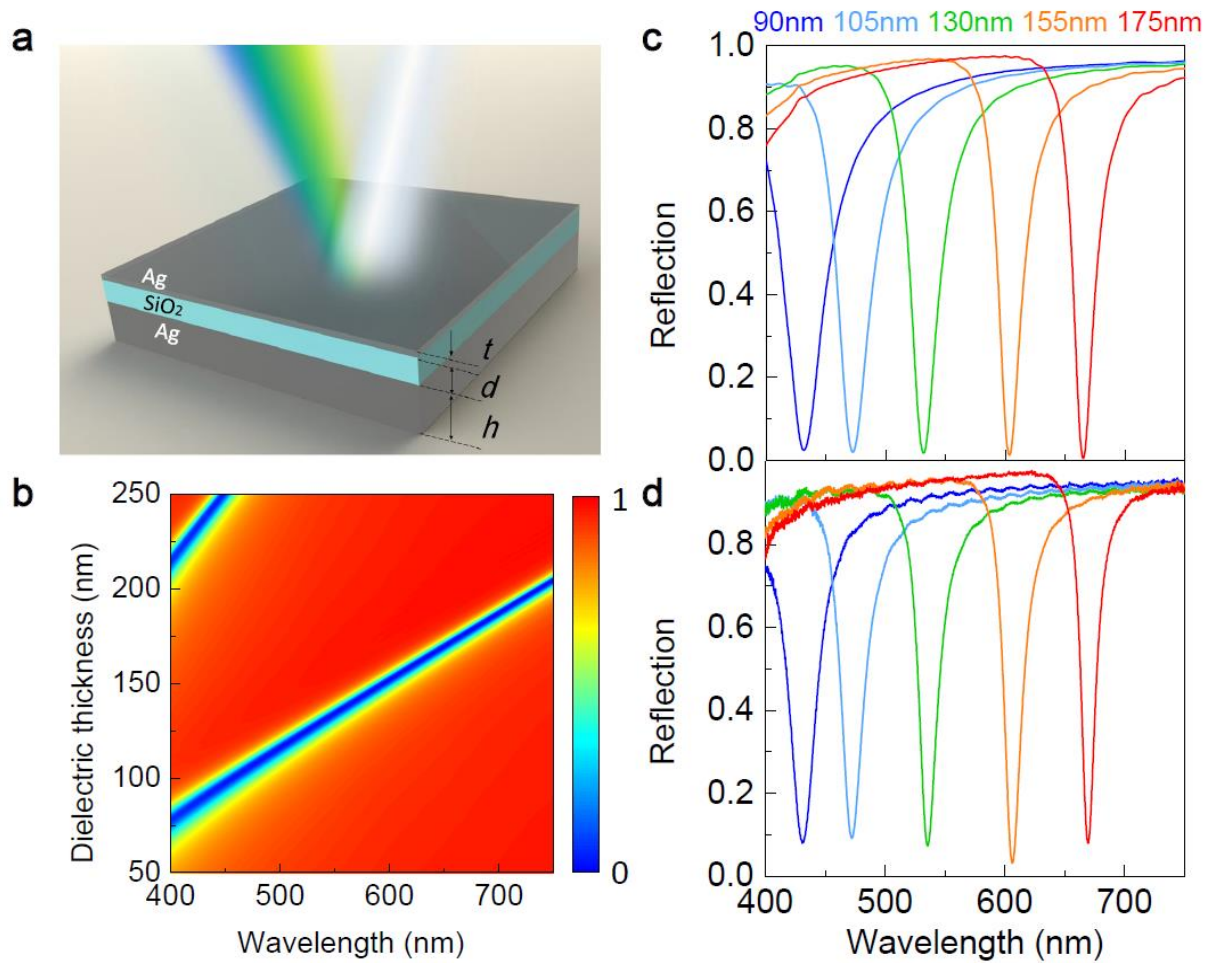

**Figure 1. Schematic configurations and reflectivity spectra of planar thin-film MIM cavity. a**, Schematic drawing of planar Ag/SiO$_2$/Ag cavity. The thickness of top and bottom metallic layers are set to be $t$ = 30 nm and $h$ = 100 m. The illumination source is normally incident onto the planar surface with arbitrary polarization. **b,** Reflection plotted as a function of wavelength and dielectric thickness $d$. **c**, Simulated and **d**, measured reflection spectra for MIM cavity with SiO$_2$ spacer thicknesses of $d$ = 90, 105, 130, 155 and 175 nm.

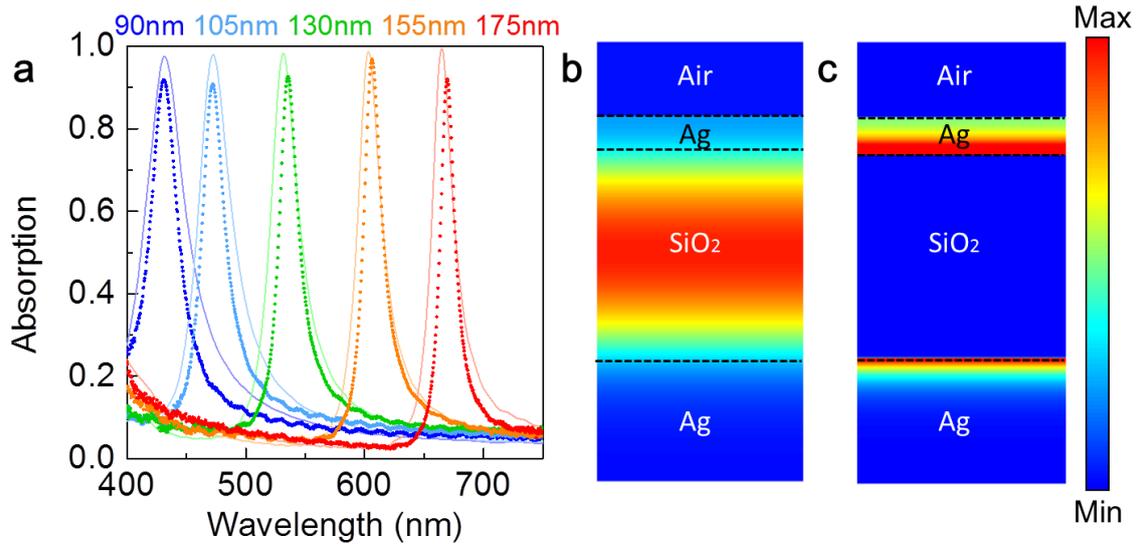

**Figure 2. Near-perfect absorption by asymmetric Fabry-Perot-type cavity. a,** Measured (dots) and simulated (solid lines) absorptivity spectra for MIM cavity with different dielectric (SiO$_2$) thickness $d$ = 90, 105, 130, 155 and 175 nm. **b,** Electric field intensity profile and **c,** absorbed power distribution for the MIM filter ($d$ = 155 nm) at the maximum absorption peak wavelength of 607 nm.

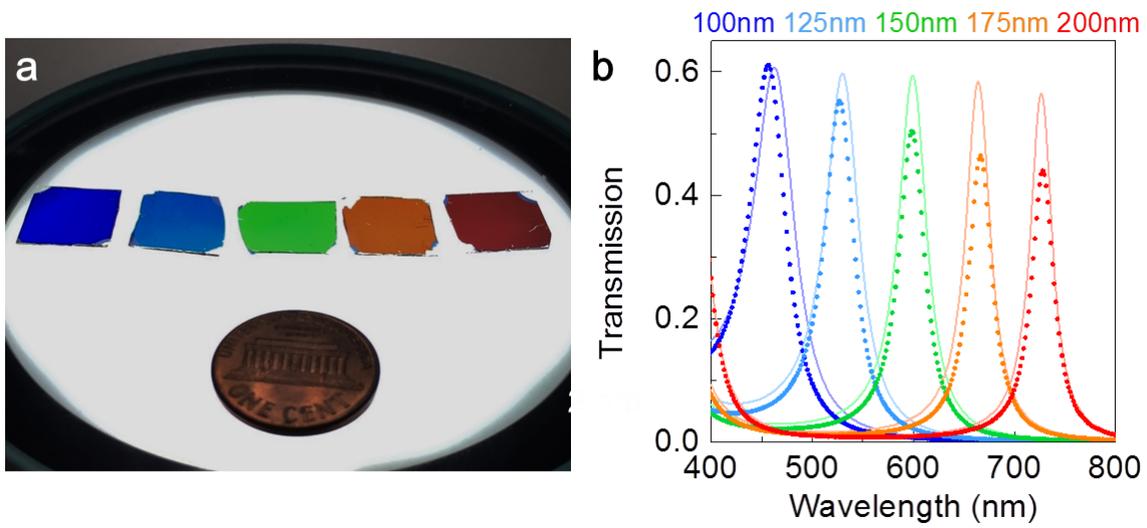

**Figure 3. Color filters. a**, Photograph of five different large-area color filters including blue, cyan, green, orange and wine. MIM samples deposited on double-side polished sapphire substrate with different oxide thickness *d*. For size comparison a United States cent is included. **b**, Measured (dots) and simulated (solid lines) transmission spectra for MIM cavities with different dielectric thickness *d* = 100, 125, 150, 175 and 200 nm.

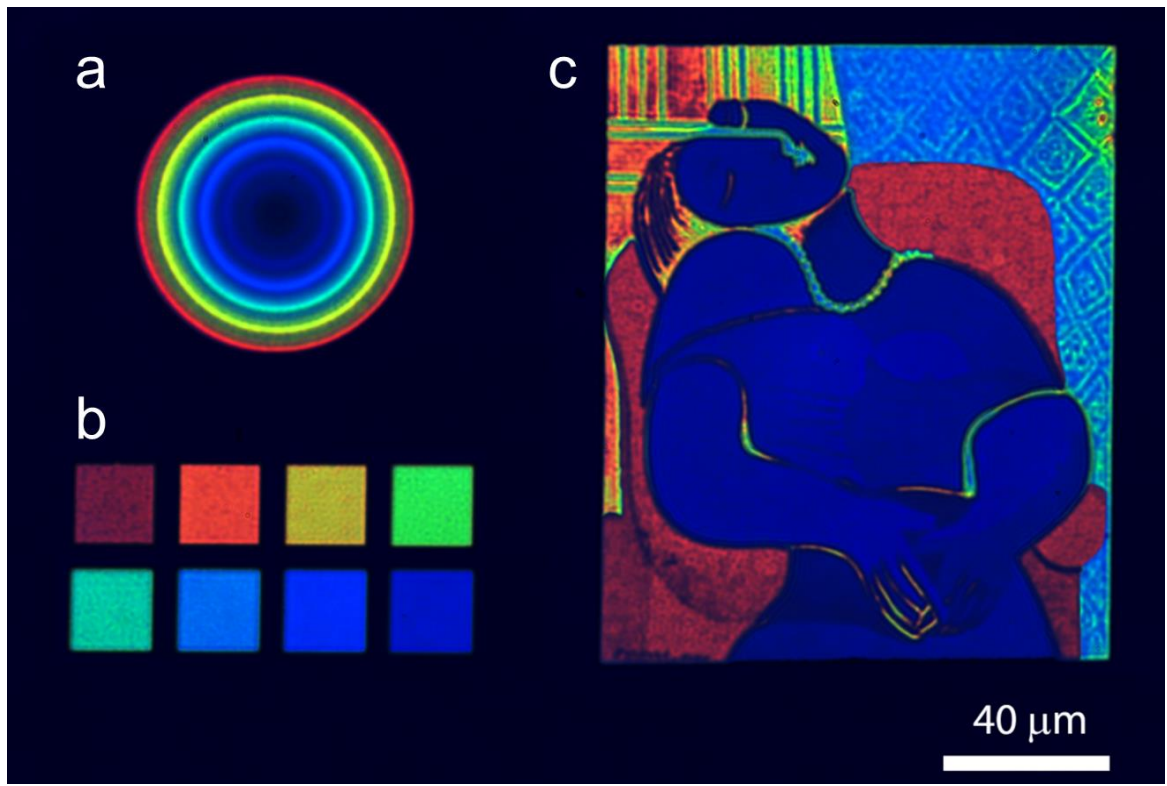

**Figure 4. Photographs of colorful images fabricated using focused ion beam (FIB) technique for patterning on the thin films with oxide thickness variation.** Among the demonstrated patterns are the rainbow circles, color palettes and the masterpiece of oil painting entitled *Le Rêve* (*The Dream)* by Pablo Picasso. The fabricated painting image measures 178K dpi of ultra-high resolution, which is ~ 500 times denser than any average smart phone display. The painting image shown here is about the size of one single pixel in an average smart phone display.

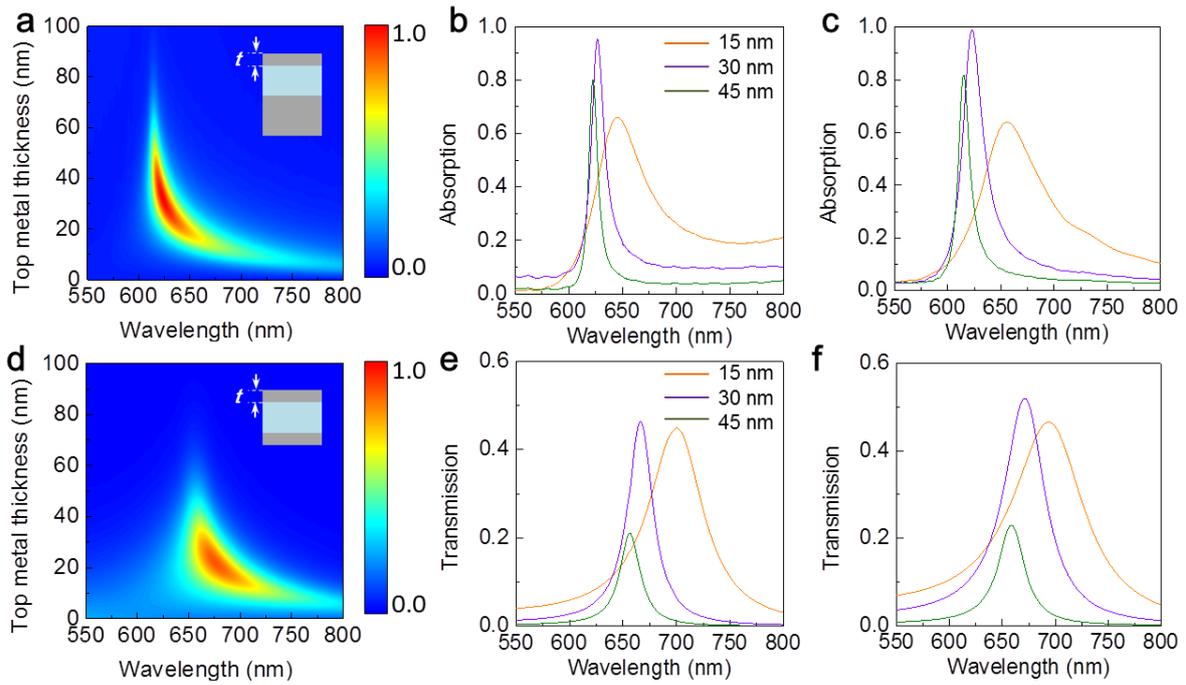

**Figure 5. Tunability of resonance peak and bandwidth in absorption and transmission spectra by top metallic film thickness. a**, Absorption evolution as a function of wavelength and top Ag thickness $t$ in MIM super absorber ($d = 160$ nm and $h = 100$ nm). **b**, Corresponding measured and **c**, simulated absorptivity spectra for MIM absorbers with different top Ag thickness $t = 15$, 30 and 45 nm. **d**, Transmission evolution as a function of wavelength and top Ag thickness $t$ in MIM transmission filter ($d = 175$ nm and $h = 30$ nm). **e**, Corresponding measured and **f**, simulated transmission spectra for MIM transmission filters with different top Ag thickness $t = 15$, 30 and 45 nm.

# Supporting Information

**Large-area, Lithography-Free Super Absorbers and Color Filters at Visible Frequencies Using Ultrathin Metallic Films**

*Zhongyang Li [†], Serkan Butun [†], and Koray Aydin**

**Performance Parameters for reflection mode and transmission mode filter**

To clearly exhibit the performance of filtering effects by thin-film coatings, we collected significant spectral features including resonance wavelength, absorption/transmission amplitude and bandwidth in experimental measurement, as displayed in Table S1. Note that the maximum absorption achieved is ~ 97% and the narrowest bandwidth measured is ~ 8.6 nm in the case of perfect absorbers (Reflection mode filters) and the maximum transmission is ~ 60% and the narrowest bandwidth measured is ~ 24.4 nm in the case of color filters (transmission filter mode).

| Reflection filter mode | | | | Transmission filter mode | | | |
|---|---|---|---|---|---|---|---|
| d (nm) | Resonance wavelength (nm) | Minimum Reflection | Bandwidth (nm) | d (nm) | Resonance wavelength (nm) | Maximum Transmission | Bandwidth (nm) |
| 90 | 431 | 0.08 | 30 | 100 | 456 | 0.60 | 44 |
| 105 | 472 | 0.09 | 19 | 125 | 527 | 0.55 | 40 |
| 130 | 535 | 0.07 | 18 | 150 | 598 | 0.50 | 33 |
| 155 | 607 | 0.03 | 17 | 175 | 667 | 0.46 | 29 |
| 175 | 669 | 0.08 | 14 | 200 | 728 | 0.44 | 27 |
| t (nm) | Resonance wavelength (nm) | Minimum Reflection | Bandwidth (nm) | t (nm) | Resonance wavelength (nm) | Maximum Transmission | Bandwidth (nm) |
| 15 | 645 | 0.36 | 57.3 | 15 | 701 | 0.44 | 63.5 |
| 30 | 627 | 0.05 | 15.1 | 30 | 667 | 0.46 | 29.0 |
| 45 | 623 | 0.20 | 8.6 | 45 | 656 | 0.21 | 24.4 |

**Table S1. Performance spectral parameters for both reflection mode and transmission mode filter.**

**Angle-dependence for FP thin film coating**

Through coating different thickness of oxide layer, transmission filters could display different color of transmission mode and it could be observed directly by human eye. It is also rather interesting to detect the angle-dependent property that the coloring illustrated by the samples are changing when observing the sample from different prospectives. As shown in Figure S2, the samples with five different oxide thickness are illuminatated by the underneath lamp and the revealing colors from samples could change when capturing the images from different prospective angles. For instance, the middle sample coated with 155 nm-thickness oxide film turns from orange to green and to light blueish when observed by human eyes from normal to different oblique angles of detection.

      Similarly with color filters, perfect absorbers based on MIM cavities are also subject to angle-dependent absorption spectra. We experimentally investigate the spectra evolution as a function of incidence angles for perfect absorber stacks. Figure S3 exhibits that resonance dips in measured reflection spectra are subject to blueshift with reduced amplitude when increasing incident angle of *P*- or *S*-polarized source from 45 ° to 80 °. In addition, the angle-sensitivity is varied for different polarization modes, e.g., *P*-polarization incident is more sensitive than *S*-polarization case owing to the stronger blue-shift in wavelength scale. The angle-dependence characteristics could suggest application in angle-sensitivity absorber and directional thermal emitters.

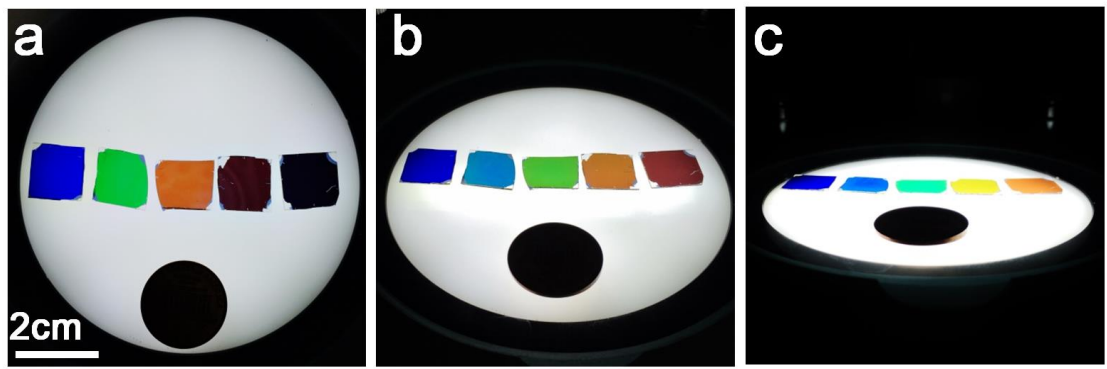

**Figure S2 Angle-dependence for transmission-mode FP-type cavity. a-c,** Color transformation when rotating the samples coated with different oxide thickness above illumination from white light source.

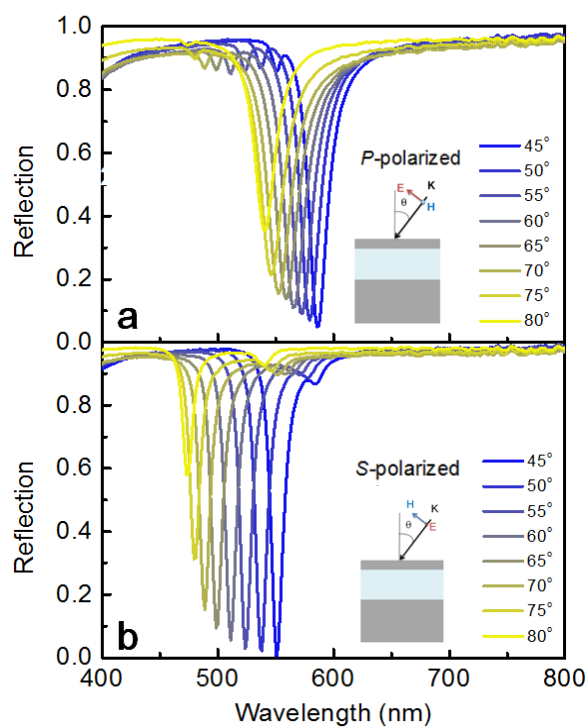

**Figure S3.** Measured reflectivity spectra for oblique incidence of **a**, *P*-polarized and **b**, *S*-polarized source as a function of different incidence angles from 45 ° to 80 °.

**Analytical modelling for the phase delay in top metal layer**

To verify the function of top metallic layer in spectra evolution shown in Figure 5 of main text, an analytical model was developed for clarifying the counter-intuitive blue-shift when increasing the top metal layer thickness. Figure S4a-S4c (Figure S4d-S4f) are respectively the experimental, simulated and modelled absorption spectra for perfect absorbers (color filters) with different thickness of top Ag film, which match well in terms of predicting the resonance positions and bandwidth. In the case of perfect absorbers, by mathematically expanding the electric and magnetic field components in each MIM region, an analytic solution could be derived for the zero-order reflection by treating the bottom refractive layer with the surface-impedance boundary conditions,[1-3] as shown in below:

$$R = |r|^2 = \left| \frac{r_1 \cdot (1-\emptyset_2) \cdot \emptyset_1 + r_0 \cdot (1-r_1^2 \cdot \emptyset_2)}{(1-r_1^2 \cdot \emptyset_2) + r_0 r_1 \cdot (1-\emptyset_2) \cdot \emptyset_1} \right|^2 \qquad \text{Eq. (1)}$$

Here $r_0$ and $r_1$ are reflection coefficient at the interfaces of air/Ag and Ag/SiO$_2$, respectively; $\emptyset_1 = e^{2ik_1 t}$, $\emptyset_2 = e^{2ik_2 d}$ are associated with the round-trip phase delay inside the top Ag and SiO$_2$ layers ($k_1$ and $k_2$ are complex wave vectors in Ag and SiO$_2$ layers, respectively). According to Equation (1), the phase shift item $\emptyset_1$ is involved in the reflectivity spectra expression, thus enabling to modify the resonance positions. Therefore, the counterintuitive blue-shift resonance is attributed to the wave phase shift within the top of metal layer.[4] For the case of color filters, we have utilized the Transfer Matrix Method (TMM) for modelling the spectra evolution when changing the top Ag thickness, as illustrated in Figure S4f.

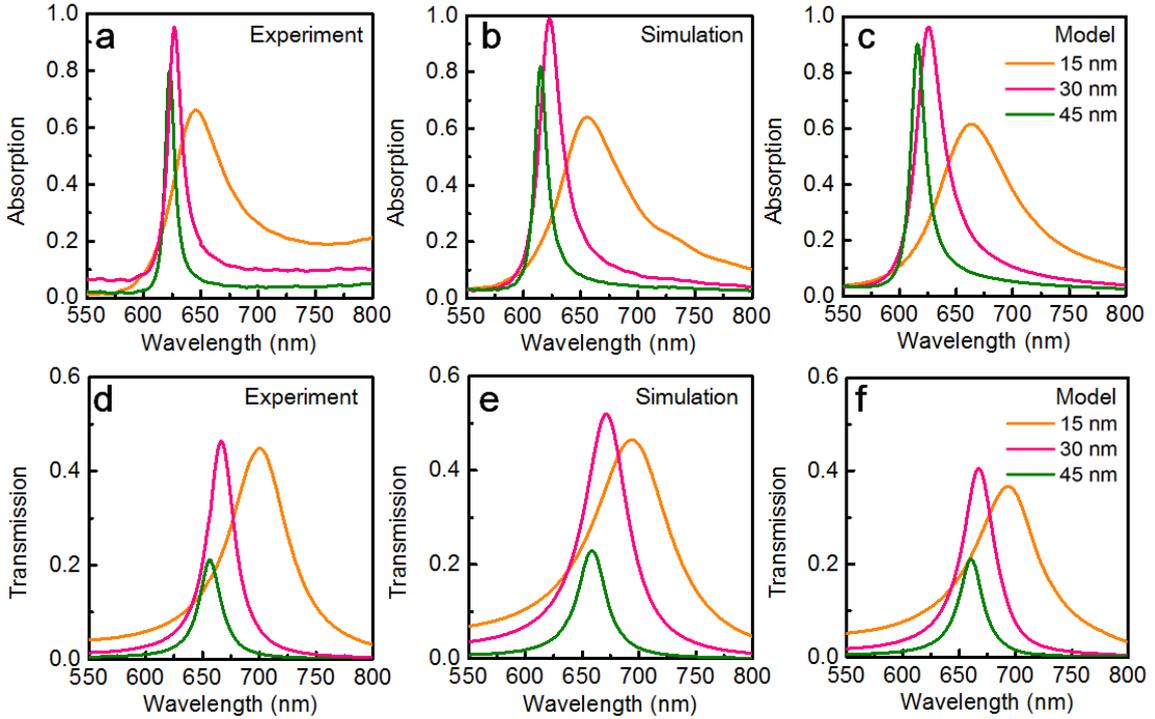

**Figure S4. Absorption spectra as a function of top layer thickness** $t$. **a, b and c** are measured, simulated and modelled absorption spectra for perfect absorbers ($d$ = 160 nm and $h$ = 100 nm) with different top Ag thickness $t$ = 15 nm, 30 nm and 45 nm. **d, e and f** are measured, simulated and modelled absorption spectra for color filters ($d$ = 175 nm and $h$ = 100 nm) with different top Ag thickness $t$ = 15 nm, 30 nm and 45 nm.

## **Metallic material comparison**

To gain more insight into the impact of metal's optical property on thin-film coating's absorption performance, four MIM stacks by Ag, Au and their hybrid constructions are numerically studied, as shown in Figure S5. Owing to the distinction between the optical properties of Ag and Au in visible domain (Figure S5a), the simulated reflection spectra demonstrates there is minor resonance shifts and amplitude variation among MIM stacks of different material constructions. For instance, the reflection resonance of Structure #4 (Au/oxide/Au) illustrates minor reduced resonance amplitude and ~ 25 nm redshift to Structure #1 (Ag/oxide/Ag). That is because Au possesses higher real part of refractive index (n) which builds longer optical cavity length, and thus larger wavelength operation. Besides, the smaller imaginary part of index (k) for Au attributes to its less lossy property and reduced absorption in MIM cavity.

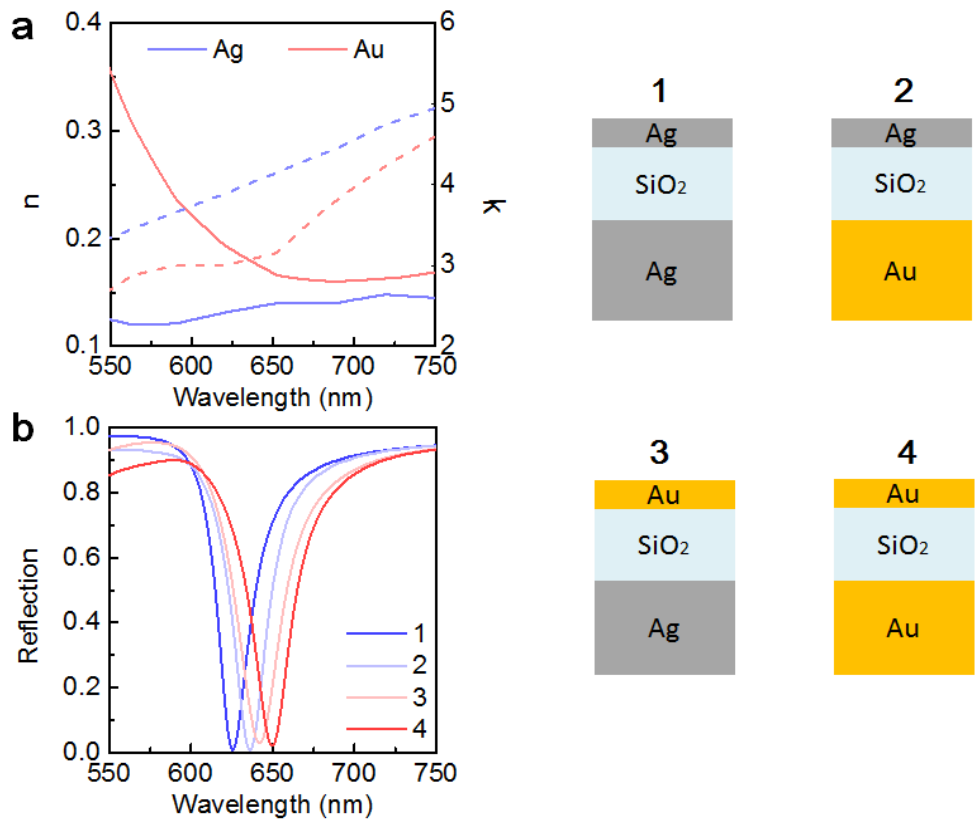

**Figure S5. a,** Real (solid curves) and imaginary (dash curves) parts of the complex refractive indices of Ag and Au. **b**, Simulated reflection spectra for MIM stacks with Ag/Au material. Structures #1-4 are respectively constructed by Ag or Au as top or bottom layer of metal.